\documentclass[runningheads]{llncs}

\usepackage{graphicx}
\usepackage{tabularx}
\usepackage{tablefootnote}
\usepackage[flushleft]{threeparttable}
\usepackage{makecell}
\usepackage{tikz-qtree}
\usepackage{subfig}
\usepackage[hyphens]{url}
\usepackage{hyperref} 
\usepackage[htt]{hyphenat} 
\usepackage{array, listings}
\usepackage{enumitem}

\newcolumntype{L}[1]{>{\raggedright\let\newline\\\arraybackslash\hspace{0pt}}m{#1}}
\makeatletter
\renewcommand\section{\@startsection{section}{1}{\z@}%
	{-8\p@ \@plus -4\p@ \@minus -4\p@}%
	{6\p@ \@plus 4\p@ \@minus 4\p@}%
	{\normalfont\large\bfseries\boldmath
		\rightskip=\z@ \@plus 8em\pretolerance=10000 }}
\renewcommand\subsection{\@startsection{subsection}{2}{\z@}%
	{-8\p@ \@plus -4\p@ \@minus -4\p@}%
	{6\p@ \@plus 4\p@ \@minus 4\p@}%
	{\normalfont\normalsize\bfseries\boldmath
		\rightskip=\z@ \@plus 8em\pretolerance=10000 }}
\renewcommand\subsubsection{\@startsection{subsubsection}{3}{\z@}%
	{-4\p@ \@plus -4\p@ \@minus -4\p@}%
	{-1.5em \@plus -0.22em \@minus -0.1em}%
	{\normalfont\normalsize\bfseries\boldmath}}
\makeatother

\begin{document}

\title{Analyzing Use of High Privileges
	on Android: \\An Empirical Case Study of
	Screenshot and Screen Recording Applications
	\thanks{This work was supported by NSFC Project 61802080.}
}

\titlerunning{Analyzing Use of High Privileges on Android}

\author{}
\authorrunning{}
\institute{}

\author{Mark H. Meng\inst{1}
	\and
	Guangdong Bai\inst{2}
	\and
	Joseph K. Liu\inst{3}
	\and
	Xiapu Luo\inst{4}
	\and 
	Yu Wang\inst{5}
}

\authorrunning{M. H. Meng et al.}

\institute{
	Institute for Infocomm Research, \\Agency for Science, Technology and Research (A*STAR), Singapore\\ 
	\and
	Griffith University, Australia\\
	\and
	Monash University, Australia\\
	\and
	Hong Kong Polytechnic University, Hong Kong S.A.R.\\
	\and
	Guangzhou University, China\\
	\email{menghs@i2r.a-star.edu.sg, 
		g.bai@griffith.edu.au, 
		joseph.liu@monash.edu, 
		csxluo@comp.polyu.edu.hk, 
		yuwang@gzhu.edu.cn}\\
}

\maketitle   
\begin{abstract}
The number of Android smartphone and tablet users has experienced a rapid growth in the past few years and it raises users' awareness on privacy and security issues of their mobile devices. 
There are lots of users rooting their Android devices for some useful functions, which are not originally provided to developers and users, such as taking screenshot and screen recording. 
However, after observing the danger of rooting devices, the developers begin to look for  non-root alternatives to implement those functions.
Android Debug Bridge (ADB) workaround is one of the best known non-root alternatives to help app gain a higher privilege on Android. 
It used to be considered as a secure practice until some cases of ADB privilege leakage have been found.
In this paper, we propose an approach to identify the potential privilege leakage in Android apps that using ADB workaround. 
We apply our approach to analyze three real-world apps that are downloaded from Google Play Store.
We then present a general methodology to conduct exploitation on those apps using ADB workaround. 
Based on our study, we suggest some mitigation techniques to help developers create their apps that not only satisfy users' needs but also protect users' privacy from similar attacks in future.
	
\keywords{Android security 
	\and application analysis 
	\and privilege escalation
	\and ADB workaround \and exploit.
}
\end{abstract}

\section{Introduction}
\label{sec:intro}
The rise of mobile devices has greatly enriched people's lives in this digital era. As the dominator of current mobile device market, Android has reserved over 77.3\% of the global smartphone market share by July of 2018 \cite{gs2018marketshare}. 

At the moment of this paper being drafted, the global number of monthly active Android devices has exceeded 2 billion \cite{ben2017google}.
The over-reliance on mobile devices makes people save all the data regardless of personal or business purpose onto their smartphones or tablets, which may lead their privacy under exposure if no proper protection has been enforced.

Android is well-known by its rich functionality and customization, but there are still some features that could not be implemented merely using the official application programming interfaces (APIs). 
Google creates a collection of permission labels to define the privilege of apps running on Android operating system (OS). Some actions like reading the content displaying on the screen, in another word taking screenshot and screen recording, are marked as \emph{signature} level permissions, hence are not allowed to be realized by common third party apps. 
However, as long as the requirement of users exists, the developers would never stop to push the boundary.
For that reason, developers are all motivated and successfully come up with two approaches to solve the permission dilemma, namely ``rooting the phone'' and ``ADB workaround''.

Rooting the devices could enable users to gain the administration privileges to do anything they want such as removing pre-installed apps, unlocking more functionalities, or changing the theme of UI. According to a statistic done by Kristijan \cite{ah2014phoneroot}, there are over 27.44\% users indicating that they have rooted their smartphones to remove redundant and useless pre-installed applications. 
There are several security issues behind the ``rooting'' because it circumvents the permission mechanism on Android system.
The good news says there is an increasing number of people who have realized the risk of rooting their devices, and have started seeking non-root approaches.
Gaining a higher privilege through \emph{Android Debug Bridge (ADB)} is one of the best known and widely used workarounds. 
Users can connect their devices to a PC via either USB or wireless network, launch the ADB and then invoke a service with system level privilege running in the background. 
After that, an application could communicate with that service, send command to it, and thereby trigger it to work for the application with system privilege. 
In this manner, that app can do the job even without APIs provided by Android. 
There are plenty of apps on Google Play Store adopting this ADB workaround to satisfy users' specific needs, including, but not be limited to, performing backup and restoration, taking screenshot, recording screen, etc. 
Those apps that use ADB workaround to achieve high privilege are very popular in recent years while Google has not yet granted corresponding permissions to developers. 

The security concern of ADB workaround has been raised up after some exploitation being successfully conducted. 
In this work, we design an approach to discover the vulnerabilities of ADB workaround.
We apply this approach to three real-world apps downloaded from Google Play Store, analyze them and eventually identify the potential privilege leakage on each of three apps. 
In addition, we conduct an exploitation on one of these three apps named ``No Root Screenshot It'' and successfully prove the existence of vulnerabilities that we have recently found. 
Based on the outcome of our exploitation, we find that all the apps found by us that uses ADB workaround to achieve privilege escalation are vulnerable to the attack through the socket channel.
Once the attacker finds a way to install the malicious application on the target device, the user privacy stored on the device will be in great risk of being stolen or leaked.
Last but not the least, we provide some advices to the developers to mitigate security risk and thereby achieve users' requirement and meanwhile protect users' data and privacy.

Therefore, this paper marks the following contributions:
\let\labelitemi\labelitemii
\begin{itemize}[noitemsep]
	\item We discuss the potential vulnerability of ADB workaround usage on Android devices by conducting our empirical case study.
	\item We propose a general approach to perform exploitation to any application using ADB workaround to achieve privilege escalation.
	\item We carry out our exploitation on a real application downloaded from Google Play Store and we prove that the ADB workaround brings with a significant security loophole.
	\item We emphasize that the security consideration during the application design and implementation is crucial to the preservation of users' privacy and hence we provide our mitigating suggestions to the developer community.
\end{itemize}

This paper is organized as follows. In the next section, we briefly introduce the security mechanism of Android, the concept of ADB workaround and related works. 
In Section~\ref{sec:approach}, we present the dataset we have collected and then we explain our approach to conduct case study. 
We also summarize a methodology to perform exploitation and we test our exploit app on actual Android devices in that section.
Section~\ref{sec:case_study} is made up of our investigation based on 3 experiments of Android applications. 
We also present our corresponding observation in each experiment.
Moreover, we provide our suggested mitigation in Section~\ref{sec:mitigation}.
Finally, we wrap up this paper in Section~\ref{sec:conclusion} with our concluding remarks.

\section{Background}
\label{sec:background}
\subsection{Privilege \& Permission on Android}

Privilege is a security attribute required for certain operations. 
In Unix-like OS, the process privileges are assigned in the principle of file system ownership. 
Its privilege mechanism is organized in shape of a flat tree where users' privilege is presented as the leaves and the superuser is described as the \emph{root} \cite{provos2003preventing}.  
Android, as a mobile operating system built based on Unix, takes advantages of the user-based privilege mechanism to identify, isolate and protect the resources used by applications. 
Every app is assigned with a unique user identification (UID), runs within the application sandbox where it only has limited permissions to access resources from the OS or other apps \cite{bugiel2012towards}. 

From the perspective of application, Android adopts the concept of ownership-based permission system from the underlying Linux kernel and develops its own access control mechanism, which is also known as the discretionary access control (DAC) \cite{chen2017analysis}. 
On Android platform, permissions are classified into several protection levels. 
Most of the Android developers are made available to the \emph{normal} level permissions and the \emph{dangerous} level permissions in their development. The \emph{normal} level permissions, such as Internet, vibration, NFC or setting alarm, are considered as having no great risk to the user's privacy or security.
It will not prompt users for consent if the usage of those permissions is properly declared in manifest during the development. 
The \emph{dangerous} level permissions indicate that the application needs access to private data or control over the device that may potentially have a negative impact to user. 
Unlike the \emph{normal} level, all the operations classified in \emph{dangerous} level will not be executed until obtaining user consent. 
In addition to aforementioned two permission levels, there are two more protection levels namely \emph{signature} level and \emph{signature or system} level defining risky permissions. 
The former is only granted to the application signed by the same encryption key with the one it declared the permission in advance. Furthermore, some \emph{signature} level permissions are not made available for third party developers and they can only be granted to a trusted party like Android development group, as Table \ref{tab:listOfSignaturePermission} shows. 
The latter could only be granted to the apps that are embedded in Android system image or signed by vendors of the system image \cite{permissionGoogle}. The grant of these two permission levels is not to be approved by users, instead, it is conducted by signature validation mechanism of Android system during installation \cite{shabtai2010google}. Many functions that users require but not provided as public APIs by Android OS, like backing up, taking screenshot and screen recording, belong to the \emph{signature} level permission. 

It is noteworthy that the DAC is only effective with the premise that all the apps are executed by an unprivileged user. Similar to other Unix-like operating systems, Android also has a number of privileged users defined in its Linux kernel, such as root, system, and radio. 
The root, for instance, is the most supreme user in Android and has full access to all apps' data. 
The Android OS does not prevent the root user or any app executed with root privilege from accessing and even modifying the resources of system or other apps \cite{sysAndKernalSec}.

\subsection{Privilege Escalation}

In order to implement the functions like backup, taking screenshot or screen recording, developers have to find a way to escalate the privilege of their apps till the \emph{signature} level or even higher. 
There are two privilege escalation approaches on Android, namely rooting and non-root workarounds.

\begin{table}[t!]
	\centering
	\caption{Some examples of \emph{signature} level permissions that are not granted to the third party developers by API level 19}
	\label{tab:listOfSignaturePermission}
	\begin{tabularx}{\linewidth}{
			>{\hsize=1.05\hsize}X
			>{\hsize=.18\hsize}X
			>{\hsize=1.77\hsize}X
		}
		\hline
		\rule{0pt}{12pt} Permission API & Level & Description \\[5pt]
		\hline
		\rule{0pt}{12pt}\texttt{BROADCAST\_SMS} & 2 & Broadcast an SMS receipt notification \\
		\texttt{CALL\_PRIVILEGED} & 1 & Initiate a call without user confirmation \\
		\texttt{CAPTURE\_AUDIO\_OUTPUT} & 19 & Capture audio output stream \\
		\texttt{CAPTURE\_VIDEO\_OUTPUT} & 19 & Capture video output stream \\
		\texttt{DELETE\_PACKAGE} & 1 & Uninstall package \\
		\texttt{DIAGNOSTIC} & 1 & Read and write the diagnostic resources \\
		\texttt{DUMP} & 1 & Retrieve state dump from system services \\
		\texttt{INSTALL\_PACKAGES} & 1 & Install packages \\
		\texttt{MODIFY\_PHONE\_STATE} & 1 & Modify phone state (e.g. power on, mmi, etc) \\
		\texttt{MOUNT\_UNMOUNT\_FILESYSTEMS} & 1 & Mount/unmount file systems or removable storage \\
		\texttt{READ\_FRAME\_BUFFER} & 1 & Access to the frame buffer data (e.g. screenshot) \\
		\texttt{READ\_LOGS} & 1 & Read system log files \\
		\texttt{REBOOT} & 1 & Reboot the system \\
		\texttt{SET\_TIME} & 8 & Set system time \\
		\texttt{WRITE\_APN\_SETTINGS} & 1 & Overwrite APN setting \\[5pt]
		\hline
	\end{tabularx}
	\vspace{-10pt}
\end{table}

\subsubsection{Rooting} ~\\
Rooting is the process of allowing users of Android devices to attain privileged control. 
Once an Android device is rooted, users can take advantage of the root privilege and arbitrarily access the system resource. 
Furthermore, users can assign specific privilege to any app installed on the rooted devices, and thereby enjoy massive personalized functionality to maximize the usage of their Android devices \cite{bishop1996unix}. 
Due to those benefits, there are plenty of users rooting their Android devices even Google officially discourage to do so \cite{chris2012case,kristiian2014over}.

Android rooting is described as a double-edged sword in the paper of \cite{zhang2015android}. It offers users with more permission and freedom to use their devices, and meanwhile, it also exposes all the data and program to the adversary and bring severe security vulnerabilities \cite{rootVsUnrooted,meng2018survey}.

\subsubsection{Non-root Alternative} ~\\
Rooting an Android device is a risky practice because it may void the warranty, brick the device and bring with numerous security vulnerabilities.
Not all Android users are willing to root their device for the exchange of additional freedom and customization.
Therefore, developers start to seek non-root alternatives to escalate privilege. 
There is an alternative approach called ADB workaround to attain high level privilege without rooting the device, and it becomes popular whilst the growth of users' concern to their device security.

Take the programmatic screenshot as an example. 
An app needs to have a \emph{signature} level permission from the system to take screenshot, which is impossible for normal developers to obtain through \emph{normal} level permission request in user interface. However, there are still two workarounds even without the permission given from Android development team: 
(1) taking screenshot on rooted devices; or 
(2) making use of a process with higher privilege to indirectly escalate the privilege of the app.
The latter approach does not require the holistic change to the Android devices like ``rooting''.
In another word, it has better security and reliability \cite{lin2014screenmilker}.

ADB is a development tool provided by Google to allow developers to debug their apps through \emph{shell} commands from their PCs. 
A process requiring \emph{signature} level permissions, such as taking screenshot, is not allowed to be implemented in app by third party developers, but could be started from an ADB shell window. That is the reason why ADB workaround could achieve a higher privilege. 
By using ADB workaround, developers could implement all methods requiring \emph{signature} level permissions, pack all of them into an executive binary that could be started on ADB and run them in the background of Android OS as a service.
As long as the service is not killed (e.g. power-off, restart), the unprivileged app could communicate with the privileged proxy to achieve the functionality which are not able to be done solely by itself.

\subsection{Access the Screen Display on Android Devices}
It is a very common demand for users to take a screenshot to save and share what is happening on her mobile device. 
Android only officially provides screenshot function to users and developers since its version 4.0. The most common way for user to capture the screen content is pressing a key combination of power key and volume down key.
However, in those earlier versions before 4.0, Android OS neither offers users a function to take screenshot, nor provides public APIs to developer to produce third party apps to do so \cite{howToTakeAScreenshot2013}. 
For those reasons, there is only one way to enable user taking screenshot on their Android device, which is privilege escalating.

Android system uses Linux OS as its kernel, and therefore it shares same approach to take screenshot with traditional Linux OS. 
In Linux system, the display output stream is managed by a software library named ``\emph{framebuffer}''. By accessing the \emph{framebuffer} library, an application or a process can obtain the display data of whole screen. In early history of Android system until 3.0, reading data from the \emph{framebuffer} is the only approach to take screenshot. 
The framebuffer approach is concise and traditional but faces some challenges. First of all, the Linux applies very strict access control to the \emph{framebuffer} library, which is borrowed by Android OS as well. 
There are only 2 user groups, \emph{root} and \emph{graphic}, being able to access data from the \emph{framebuffer} on Android platform. Moreover, nowadays Android apps become complex and sometimes using multiple \emph{framebuffers} to form an overlaid display. Reading \emph{framebuffer} is very likely no longer capable to obtain the entire screen display.

Starting from version 4.0. Google introduces an interface specially for taking screenshot called \emph{SurfaceFliger}, together with a permission called \texttt{READ\_FRAME\_BUFFER} to invoke that interface. Nonetheless, Android remains its strict access control policy to the new API. Only the apps running with \emph{system} or \emph{graphics} user group are eligible to use such API to take screenshot -- which is impossible for normal third party apps to achieve.

The developer community can always find a solution although there are number of restrictions to achieve screenshot. ``\emph{ddms}'' is the most popular approach which adopts the idea of ADB workaround to eliminate the privilege restriction of screenshot taking within a third party app. 
\emph{ddms} refers to the \emph{Dalvik Debug Monitor Server}, which is a debugging tool brought with Android SDK and is also integrated into the official Android development software called Android Studio. 
By accessing the \emph{ddms}, user can make use of a third party process to send commands to the \emph{framebuffer} service through the ADB channel. Unlike the third party app itself, an ADB session is given the shell user permission, at which all processes launched in an active ADB session are eligible to be assigned with privileges of \emph{graphic} user group. Hence the \emph{ddms} approach could achieve the screenshot functionality without needs to gain a higher privilege \cite{ferrill2011navigating}. 

\subsection{Related Work}
\label{sec:related_work}

There are some previous studies unveiling the security risk of ADB workaround despite it is considered much safer than device rooting. Security concern of ADB workaround mainly comes from the difference between roles of proxy and application on Android OS. In this project, these risks could be summarized into two types: \vspace{2pt} 

(1) whether other apps could obtain control to the opening proxy by sending commands; and 

(2) whether the communication between app and proxy is properly protected if the scenario of (1) is possible to happen.\vspace{2pt} 

The description of the first kind of security concern could be found in the paper written by Lin, et al. \cite{lin2014screenmilker}. 
The communication channel between the application and its ADB proxy relies on network sockets without any protection enforced. 
For that reason, once an ADB proxy has been activated, any application has the privilege to communicate with it and even request service from it at any time without restrictions. 
This vulnerability gives attackers a chance to analyze the protocol of such communication and build a malicious application to request service from ADB proxy exactly as same as what genuine application does.

Some developers have realized the fact that the communication channel between the application and ADB proxy may be risky, and therefore implemented some authentication routines to strengthen security. However in the paper of Bai, et al, it was proved that such authentication was ineffective as long as the reverse engineering and analysis being feasible on given application. What developer can do to secure the communication is only applying some basic authentication since there is no way to enforce strong protection onto the socket network. That authentication is usually very weak in front of analysis \cite{bai2015all}.
Some application like ``\emph{Helium}'', a backup/restore application mentioned in the paper written by Bai, et al., has been found using protection during the communication between application and ADB proxy. ADB proxy requests a password that sent out from a specific process to provide service. Unfortunately, vulnerability was found in the protocol of password distribution.
The password generated each time when ADB proxy being activated, and it is independent of app's life cycle. In this way, the proxy has to find a place that readable by apps executed with user group privilege, save the password into a file and waiting for app to read from it. This life cycle inconsistency makes adversary possible to find the current using password and thereby exploit the Android device by carrying out a replay attack to the ADB proxy.

\section{Approach}
\label{sec:approach}

The Android app using ADB workaround is usually a combination of a normal application with restricted permissions, and a proxy started by ADB which has \emph{signature} level permissions. 
In Android, most of apps communicate with proxies through the socket channel, which has no strong protection and generic access control. A malicious app could easily obtain the control of proxy if it knows the protocol of communication between app and itself. The security concern arises if the proxy interface is not well protected against the third party access.
Some apps implement password authentication into the protocol to strengthen protection to the proxy. However, due to the inconsistency of app and proxy's life cycles, there usually be a mechanism to temporarily save the password. By this means, a malicious app could still have chance to obtain the password if proper analysis has been done. Therefore, an ineffective or insecure mechanism of password authentication constitutes another potential security concern.

In this work, we raise our hypothesis that all the apps using ADB workaround to attain a higher privilege are vulnerable to the attack. 
To prove that hypothesis, we collect a number of Android apps from Google Play Store. By filtering out those apps that do not adopt ADB workaround, we conduct a series of analysis in~\ref{subsec:assessment} to find out their mechanisms to achieve privilege escalation.
\emph{Static analysis} is the first step of application assessment, which will be conducted on both the proxy activation program running on the PC and the app itself. 
Static analysis helps us locate the involved classes for the proxy communication and thereby gain the knowledge of the overall procedure. 
\emph{Dynamic analysis}, on the other hand, is capable of elaborating the runtime behavior of the target app and exposing the potential error and vulnerability.
Dynamic analysis, such as hooking, is a good complement of the static analysis for our app assessment especially in case of strong obfuscation has been enforced. 
The protocol between the proxy and the app is supposed to be completely discovered after the static analysis and dynamic analysis.
For the purpose to conduct exploitation and thereby prove our hypothesis, we may also need to conduct authentication analysis to bypass the limited security mechanism applied in the target app.
With all the key information gathered, we shall proceed to exploitation design, which will be introduced in~\ref{subsec:exploit}.

\subsection{Data Set}

We collect a batch of 13 screenshot apps and 2 screen recording apps from Google Play Store, with the criteria that the app must be compatible with earlier Android versions that do not have official support of screenshot functions. 
We install those apps on a Nexus 7 device installed with Android 4.4, followed by reading their official user instructions and observing the functionality of each of them.
Those 15 apps, as shown in Table~\ref{tab:listOfApps}, covers all well-known approaches to take screenshots or screen recording on Android devices. 
There are 4 apps using ADB workaround approach to enable users to take screenshots or recording without needs to root their devices in advance. 6 other screenshot apps achieve screenshot by asking user to press key combination (e.g. power key and volume down key). Those apps essentially do not contain screenshot implementation, instead they detect the device configuration and then display the screenshot instruction if either the corresponding manufacture has official built-in function, or the Android system version installed is 4.0 or later \cite{chris2012howto}, to help users to achieve screenshot functionality -- in another word, those 6 apps are more like an assistant to guide users to take and manage screenshot pictures.
Moreover, there are 5 more apps explicitly declaring that they are not working on devices without being rooted. 
In this paper, we only focus our study on those apps that using ADB workaround.

\begin{table}[!t]
	\centering
	\caption{List of screenshot and screen recording apps found on Google Play Store}
	\vspace{5pt}
	\label{tab:listOfApps}
	\begin{threeparttable}
		\begin{tabularx}{\linewidth}{
				>{\hsize=0.2\hsize}X
				>{\hsize=3.35\hsize}X
				>{\centering\hsize=0.7\hsize}X
				>{\centering\hsize=0.6\hsize}X
				>{\centering\hsize=0.8\hsize}X
				>{\hsize=0.35\hsize}X
			}
			\hline
			\rule{0pt}{16pt}\makecell[lc]{\#} & \makecell[l]{App Name \& Identity Package Name} & \makecell{Root\\Required} & \makecell{App\\Type\tnote{1}} &  \makecell{Unrooted\\Approach} & \makecell{Size} \\[5pt] \hline
			1 & \rule{0pt}{16pt}\makecell[l]{Screen Capture - Sigourney\\\textit{com.mobilescreen.capture}} & No & S & Hardkey  & 5.2M \\ \hline
			2 & \rule{0pt}{16pt}\makecell[l]{Screenshot Easy\\\textit{com.icecoldapps.screenshoteasy}} & No & S & Hardkey  & 5.2M \\ \hline
			3 & \rule{0pt}{16pt}\makecell[l]{Screenshot Ultimate\\\textit{com.icecoldapps.screenshotultimate}} & No & S & ADB & 3.2M \\ \hline
			4 & \rule{0pt}{16pt}\makecell[l]{Screenshot Capture\\\textit{com.tools.screenshot}} & No & S & Hardkey  & 3.1M \\ \hline
			5 & \rule{0pt}{16pt}\makecell[l]{NoRoot Screenshot Lite\\\textit{com.mobikasa.screenshot.lite}} & Yes & S & N.A.\tnote{2} & 545k \\ \hline
			6 & \rule{0pt}{16pt}\makecell[l]{Screenshot and Draw\\\textit{com.conditiondelta.screenshotanddraw.trial}} & Yes & S & N.A. & 1.1M \\ \hline
			7 & \rule{0pt}{16pt}\makecell[l]{Screenshot\\\textit{com.enlightment.screenshot}} & No & S & Hardkey  & 2.4M \\ \hline
			8 & \rule{0pt}{16pt}\makecell[l]{Screenshot\\\textit{com.geekslab.screenshot}} & No & S & Hardkey  & 1.2M \\ \hline
			9 & \rule{0pt}{16pt}\makecell[l]{Screenshot\\\textit{com.icondice.screenshot}} & No & S & Hardkey\tnote{3} & 4.86M \\ \hline
			10 & \rule{0pt}{16pt}\makecell[l]{Screenshot\\\textit{com.geeksoft.screenshot}} & Yes & S & N.A. & 2.3M \\ \hline
			11 & \rule{0pt}{16pt}\makecell[l]{Screenshot ER Demo\\\textit{fahrbot.apps.screen.demo}} & Yes & S & N.A. & 3.2M \\ \hline
			12 & \rule{0pt}{16pt}\makecell[l]{No Root Screenshot It \\\textit{com.edwardkim.android.screenshotitfullnoroot}} & No & S & ADB & 838k \\ \hline
			13 & \rule{0pt}{16pt}\makecell[l]{Screenshot It\\\textit{com.edwardkim.android.screenshotitfull}} & Yes & S & N.A. & 840k \\ \hline
			14 & \rule{0pt}{16pt}\makecell[l]{FREE screen recorder NO ROOT\\\textit{uk.org.invisibility.recordablefree}} & No & R & ADB & 7.5M \\ \hline
			15 & \rule{0pt}{16pt}\makecell[l]{Mobizen Screen Recorder -- Record, Capture, \\Edit 3.1.0\qquad\textit{com.rsupport.mvagent}} & No & R & ADB & 19.9M \\ \hline
		\end{tabularx}
		\begin{tablenotes}
			\item[1] `S' stands for screenshot app and `R' stands for screen recording apps.
			\item[2] N.A.indicates that application only work on rooted devices.
			\item[3] Only compatible with devices made by some fixed manufactures.
		\end{tablenotes}
	\end{threeparttable}
	\vspace{-15pt}
\end{table}

\subsection{Application Assessment}
\label{subsec:assessment}

Executing the application on an Android device by following the user instruction is obviously not sufficient for the purpose of application assessment. 
A complete application assessment is composed of static analysis and dynamic analysis.
Fig.~\ref{fig:evaluationMethods} illustrates some common approaches to conduct application assessment on Android platform. 
The static analysis is to find a rough picture of the functionality of an Android application by analyzing source code, binary or other supporting materials such as the manifest file.
While the dynamic analysis makes use of the findings from static analysis, and consequently unveils the runtime behavior of the target application \cite{batyuk2011using}.
In this work, our approach to analyze the app and find the vulnerabilities of ADB workaround is initiated based on two potential concerns that we have mentioned above.
We summarized our approach into four step:

\begin{figure}[t!]
	\centering
	\resizebox {0.8\columnwidth} {!} {
		\begin{tikzpicture}
		\tikzstyle{every node}=[font=\small]
		\Tree[ .{Analysis approaches} [ .{Static analysis}
		[ .{Script analysis} ][ .{Decompilation}
		[ .{\makecell[c]{smali/Java code\\analysis}} ]
		[ .{\makecell[c]{Disassembly\\inspection}} ]]]
		[ .{Dynamic analysis} [ .{\makecell[c]{Runtime logs\\analysis}} ]
		[ .{\makecell[c]{Hooking method\\invocation}} ]]]  
		\end{tikzpicture}
	}
	\caption{Approach of app analysis}
	\vspace{-18pt}
	\label{fig:evaluationMethods}
\end{figure}

(1) \textbf{Analysis on the proxy activation.} This analysis could be done on reading proxy activation script if exists. 
The script is usually a batch file or bash script, which depends on the OS environment, i.e. the Windows or Linux, to be run with. 
Some apps do not provide script file to the user for Windows OS, for instead, a desktop application with graphic user interface (GUI) is provided to achieve better user experience. 
In this circumstance, the Linux version of activation package is recommended to be download because the script file is more widely used on Linux OS. 
A script file could disclose some details of the protocol of communication between app and proxy, such as the name of service proxy, the native executive file of proxy if any, and how the service being activated. Besides the analysis onto the script file, the name, process ID and permission group of service proxy running in the background could also be found by typing command ``\texttt{adb shell ps}'' in ADB through USB to the device. 
Moreover, the port opened for the communication between service proxy and app could be found in similar way by typing ADB command ``\texttt{adb shell netstat}'' to retrieve all active network usage on the device. 
However, in this step, the pairing of process and specific port listening may not be able to be observed if multiple proxies had been activated.

(2) \textbf{Analysis on the apk file.} Reverse engineering, such as apk decompilation, is involved in this step. 
Once the service proxy and port number have been identified, the next step is to discover the implementation of the communication between proxy and app in apk file. 
The apk file could be unpackaged and then decompiled into smali/Java source code by using tools like \emph{Apktool} or \emph{dex2jar}. 
The smali/Java code has supreme readability which may help us to look through different classes to locate the code of protocol's implementation. 
In fact, the decompilation analysis may not always to be proved as a smooth and easy process because a large amount of developers obfuscate their code before releasing the apk files to the app store \cite{wang2017changed}. 
There are a number of Android obfuscation tools available on the Internet that facilitate developers to obfuscate their apps to preserve copyright and intellectual property \cite{zhang2018progressive}.
In this situation, the disassembly will be helpful and a supplement to the smali/Java code reading. Reading assembly code could help us recognize the constant strings and numbers defined within same class. 

(3) \textbf{Dynamic analysis.} Only reading the script and source code may not be sufficient to sketch out the entire protocol between proxy and app. The objective of dynamic analysis is to find both control flow and data flow occurred when the app interacts with service proxy. Reading logs through \texttt{logcat} is a simple but effective way to gain a brief understanding to the protocol. However, hooking by \emph{Xposed} framework will be one of the best solutions to complete the analysis when the source code has been enforced with strict obfuscation or an authentication has been applied onto the socket channel between app and proxy server. Hooking method could be enforced onto the key methods in the class that takes responsibility to the communication between app and proxy, then sniff and extract the arguments passed in and return value through the system logs. 
According to the case studies in this work, the methods to be hooked are mostly used to handle the action trigger (e.g. \texttt{takeScreenshot}) and socket channel I/O (e.g. \texttt{write}). 
Hooking on the prior method(s) by printing logs could show us the control flow of the protocol, and hooking on the latter method(s) by extracting arguments' value could help us understand the data flow between service proxy and app itself. 
By now with both control flow and data flow confirmed, the communication protocol has been unveiled.

(4) \textbf{Authentication analysis.} There is very likely an authentication process if any series of numbers or a random string to be found in the data flow of the protocol. 
In that case, it is encouraged to clarify if the password is a constant string or dynamically generated. 
For the dynamically generated password, the password issuing should be solely performed by either the proxy or the app itself. 
The password is generally stored at somewhere that both the app and proxy have permission to read.

Once these four steps listed above have been fully understood and conducted, attackers are theoretically able to exploit an app that uses ADB workaround in a programmatic manner. 
We will perform an empirical case study of 3 apps that uses ADB workaround and we will present our findings in next section. 

\subsection{Exploitation}
\label{subsec:exploit}

Theoretically, the user privacy displayed on device could be unperceivably compromised at arbitrary time if there is a malicious app installed on that device where the proxy of original app running in the background.
In this subsection, we present our methodology to conduct the exploitation. 
Furthermore, we also select one of those 3 apps that are mentioned in the case study as the target to carry out a real exploitation, and we depict the implementation details as well as the exploitation outcome.

Based on our analysis on the apps collected for this work, we find all the apps that uses ADB workaround are vulnerable by attacking through socket channel and thereby obtain screenshot or screen recording of victim's device. 
Moreover, we find there is a large group of Android apps using socket channel communication aided by ADB workaround to achieve privilege escalation. 
For those reasons, we summarize the exploitation methodology into a set of technical processes and we extend our focus scope to all apps that use ADB workaround to achieve privileged functionality.

\begin{figure}[t!]
	\vspace{-10pt}
	\begin{center}
		\includegraphics[width=0.9\linewidth]{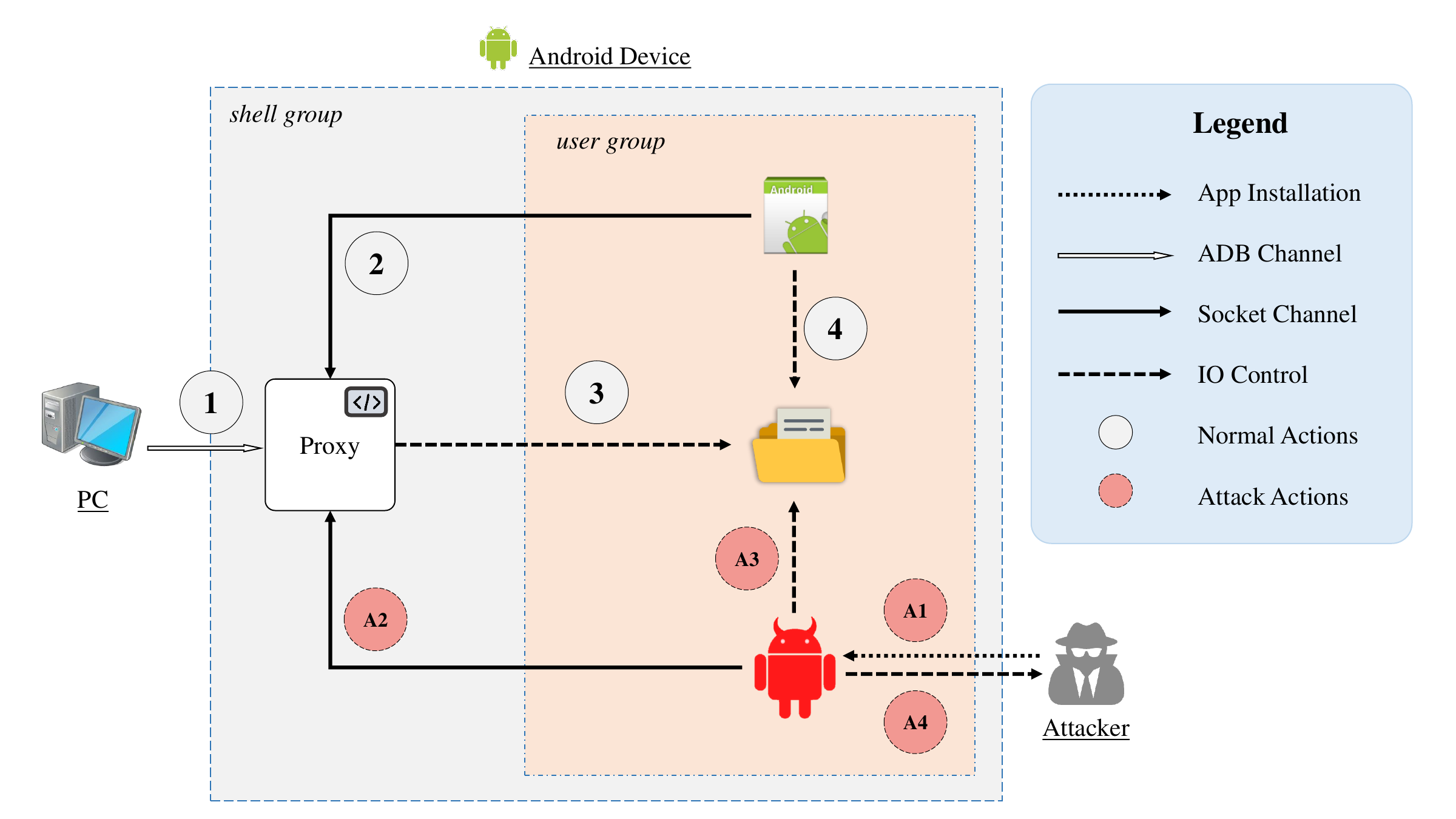}
	\end{center}
	\vspace{-15pt}
	\caption{Process to conduct ADB workaround exploitation}
	\label{fig:general-attack}
	\vspace{-15pt}
\end{figure}

As shown in Fig.~\ref{fig:general-attack}, the exploitation is achieved by replay attack initiated by a malicious app, which follows the same protocol as the original app but without compliance with users' control over their devices. 
It could carry out theft of user's privacy at any time as long as the proxy is running in the background.
Generally a successful exploitation is constituted by 4 key steps, which are:\vspace{5pt} 

$(A1)$ the attacker finds a way to install the exploitation app on the victim's device, where the benign app has also been installed on.

$(A2)$ the malicious app identifies the proxy and then conduct a replay attack;

$(A3)$ the malicious app gains access to the specific file directory where the output media files locate; and

$(A4)$ the malicious app finds a way to transmit the stolen data to the attacker.\vspace{5pt} 

In this paper, we introduce our exploitation conducted to the app II and then we present the outcome of exploitation.

We implemented an app named ``\emph{exploitNoRootScreenshotIt}'' simulating the malicious exploitation of the app named ``\emph{No Root Screenshot It}'' (app II in following case study) for the demonstration purpose.\footnote{The source code of our exploitation could be downloaded from http://mark-h-meng.github.io/attachments/analysing-use-of-high-privileges/source\_code\_folder.zip}. 
In that exploitation app, there are in total 4 messages being organized into 2 batches and sent out to the \texttt{localhost} on port \texttt{6003} through the socket channel. 
The first 2 messages are used for the configuration purpose.
Once the acknowledgment of first batch messages has been received from the proxy, which is ``\texttt{screenshotService}'' running in the background, the last 2 messages are sent out as screenshot taking commands.

The screenshot obtained is converted to a \emph{bmp} file under the sub-directory named ``\texttt{temp}''\footnote{The full directory path is /data/data/com.edwardkim.android.screenshotitfullnoro ot/temp}. The access permission of that folder was set as read-only to the user group. Therefore, once the screenshot has been taken by the proxy, the exploitation app could access to the newly captured screenshot located in the ``\texttt{temp}'' folder and make a copy to the target location such as folder under external storage ``\texttt{{/sdcard/hack\_screenshots/}}''.
The screenshot image is renamed according to the capture time to avoid being overwritten and facilitate maintenance at the same time. 
As the result, our exploit app has been successfully tested on 2 devices in our lab (a \emph{Nexus 7} with Android 4.4 installed, and a \emph{Xiaomi Rednote 3G} with Android 4.2 installed). 
This exploitation could even been further designed and programmed to take screenshot automatically with specific frequency without any notice of user, hence the user's privacy could be consequently exposed to the attacker. 

\section{Empirical Case Study}
\label{sec:case_study}
In this section, we perform our case studies on 3 apps that use ADB workaround to achieve screenshot function.
Firstly, we analysis the app titled as \emph{Screenshot Ultimate} developed by ``icecoldapps'' and we note it as app I. Then we study the app named \emph{No Root Screenshot It} developed by ``edwardkim'', which is represented by app II. After that, we conduct our analysis on the third app called \emph{FREE screen recorder NO ROOT}, which is produced by ``Invisibility Ltd'' and noted as app III.

\subsection{App I -- Screenshot Ultimate.}

``\emph{Screenshot Ultimate}'' is a typical screenshot app that does not require a rooted device. It supports screenshot taking through ADB workaround. 
However, that usage is veiled since too obvious instruction may lead to a ban from Google. 
The ADB workaround is mentioned in a paragraph of ``Help'' instruction, and the URL to download the script and other necessary files are given in another place and could only be found on the screen display within the app.
The developer has provided detailed step-by-step instruction and troubleshooting notes.

\subsubsection{Analysis on the proxy activation}~\\\vspace{-8pt}

The native executable file, named ``\texttt{screenshotultimatenative1}'', and scri-pts for both Linux and Windows OS could be downloaded as a zipped file from the URL given in the help instruction. 
After reading through the script file, we found the execution of script pushed that native executable to the file directory of the application in the device, then configured another native executable named \texttt{absel} located in the application file directory to user executable mode, and finally launched both native execution files to make them run in the background.
We summarize the flow of service activation and show it in Fig.~\ref{fig:serviceActivationIcecode}. 
With the process name of service running in the background, we can analyze the apk file and unveil the protocol of screenshot taking process between app and that service.

\subsubsection{Analysis on the apk file}~\\\vspace{-8pt}

The reverse engineering tool ``dex2jar'' is used to decompile the apk file to the jar format. Then further Java decompilation has been done by ``JD-GUI''. 
Unfortunately, the class organization of the source code obtained from the decompilation of ``Screenshot Ultimate'' is not quite readable because the obfuscation is believed to be applied. 
Some core methods which control the logic flow of screenshot taking are missing. 
Clues could only be found by analyzing package structure, libraries imported and source code from the remaining classes.  

Obfuscation cannot perfectly hide everything in the decompiled source code. 
After carefully reading through the source code of ``\emph{Screenshot Ultimate}'', we find some clues to shape the mechanism of screenshot taking. 
For example, there is an address in Android OS partition, which is in form of a string variable with the value as ``\texttt{/system/bin/fbread}'', appearing more than once in the obfuscated classes.
Its occurrence suggests that this app is very likely taking screenshot by reading image data from \emph{framebuffer}, which is commonly expressed as short writing ``fb'' in Android development. 
Reading framebuffer to take screenshot is usually achieved by a library called \emph{Android screenshot library} (known as ``ASL'')\footnote{Android screenshot library is available at \url{https://code.google.com/archive/p/android-screenshot-library/downloads}}. 
We downloaded the ASL from Android open source repository and compared the checksum value with the native executive that we downloaded from the URL given by the app developer. 
The comparison result reveals that 2 executive files are exactly equivalent, which proves our hypothesis that the app I makes use of ASL to read framebuffer and thereby capture the screen display.

\begin{figure}[!tbp]
	\centering
	\begin{minipage}[t][][b]{0.42\textwidth}
		\centering
		\includegraphics[width=0.8\linewidth]{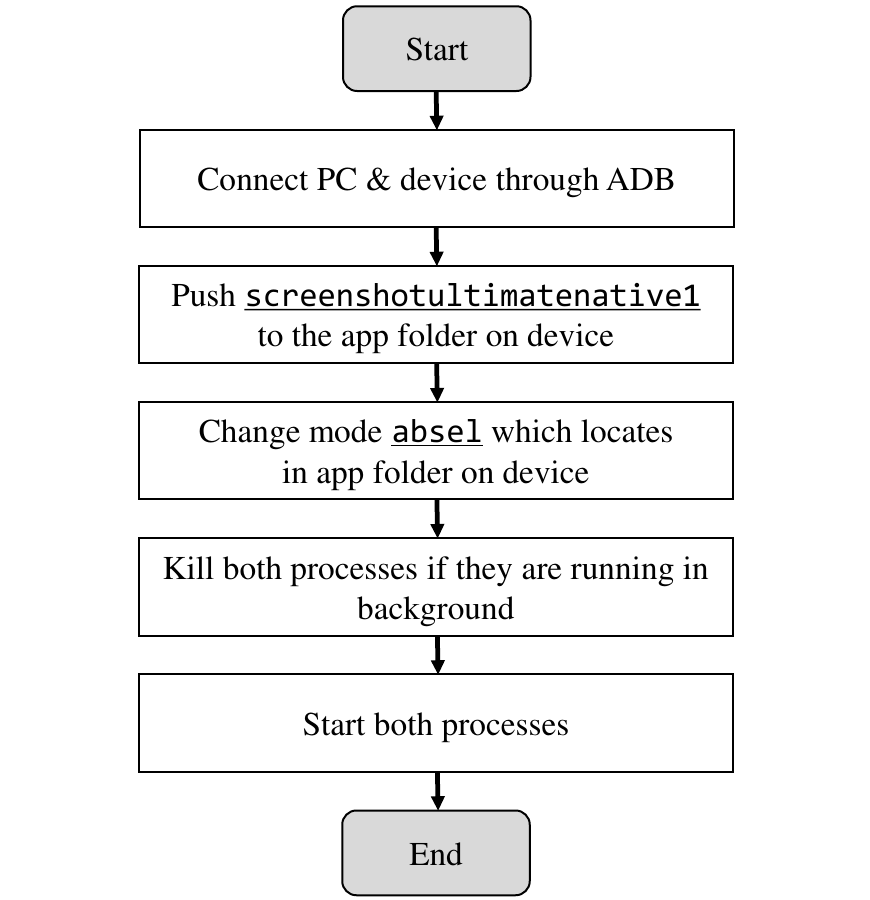}
		\caption{Proxy activation of app I}
		\label{fig:serviceActivationIcecode}
	\end{minipage}
	\hfill
	\begin{minipage}[t][][b]{0.56\textwidth}
		\centering
		\vspace{-8px}
		\includegraphics[width=1\textwidth]{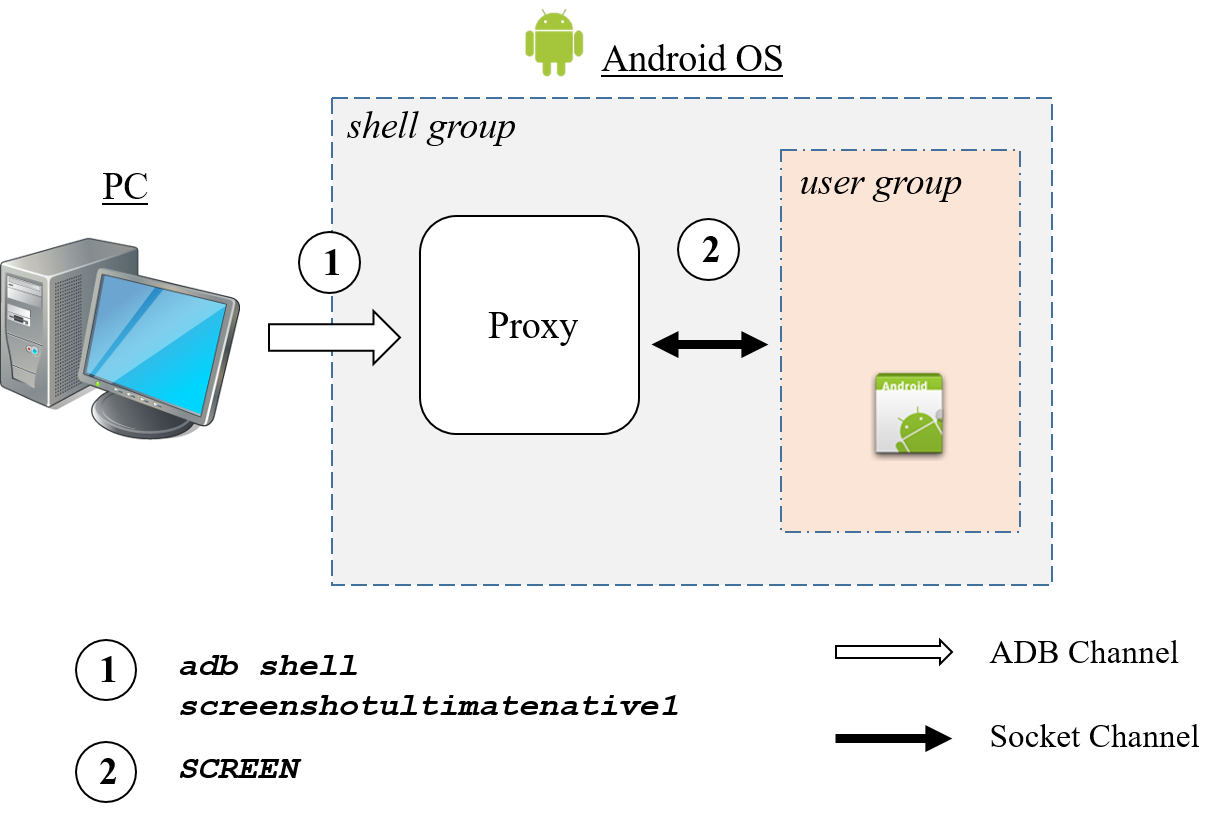}
		\caption{Process of taking screenshot on app I}
		\label{fig:processDiagramIcecold}
	\end{minipage}
	\vspace{-15px}
\end{figure}

The ASL enables Android developer to write screenshot app without root requirement. 
Once the user follows the instruction and executed the native executive file by running the given scripts, proxy with shell permission could help user take screenshot which the application has no privilege to do so. Take ``Screenshot Ultimate'' as example, user could just click the ``Screenshot'' button at the moment that user wants to take screenshot of his/her device, then the app send the screenshot command to the proxy running in background via socket channel, following by the proxy as a process named ``\texttt{screenshotultimatenative1}'' reading the current hardware framebuffer, converting to the image format and saving to the specific location.
Furthermore, we find the command that the app I sends to the background service through socket channel. The message is a constant string with a value as ``\texttt{SCREEN}''.
The communication is carried out by a plaintext messaging mechanism through a fixed port number, which is obviously not secure at all. 
In the end, we sketch out the protocol of communication between app and proxy and place it into a bigger scale of the entire life cycle of the app I. We present the process diagram of app I in Fig.~\ref{fig:processDiagramIcecold}.

\subsection{App II -- ``No Root Screentshot It''}

Unlike the ``\emph{Screenshot Ultimate}'', the app II ``\emph{No Root Screenshot It}'' has additional security feature and protection enforcement being implemented during the development.
The obfuscation has been conducted onto both service activator and apk file. 
Meanwhile, the communication channel between app and proxy has also been protected by using some identification trick like a password.

\subsubsection{Analysis on the proxy activation}~\\\vspace{-8pt}

Instead of simply running a batch script, the service activation of app II is performed by executing a \emph{.Net} application named ``Screenshot It Enabler''. 
Therefore the decompilation of \emph{.Net} application is involved in the static analysis of app II.
Moreover, the script file was not found in the enabler's package, which means it has been packaged into the apk and the purpose of the enabler is just to run the ``shell'' command to execute it. 
A \emph{.Net} decompilation tool named ``JetBrains dotPeek''\footnote{dotPeek is available on {https://www.jetbrains.com/decompiler/}} has been used to conduct the reverse engineering of the activation tool. 
Even though the enabler application has been obfuscated, some variables and C\# code logics could still be recovered after the decompilation. The scripts to enable the proxy has been unveiled by observing the C\# code from the decompilation result. 
We noticed there is a string ``screenshot'' that occurs in the decompiled C\# code as one of the argument while launching ADB service. For that reason, we believe that there is a script file named ``screenshot'' being executed during the service activation. 
Then the script file's location could be easily found by browsing the file manager on a rooted phone, or by decompiling the apk file and searching the file name. 

\subsubsection{Analysis on the apk file}~\\\vspace{-8pt}

After clarifying the ADB communication to activate the proxy, the following step focuses on discovering the communication between app and the proxy, thereby obtain the commands to control proxy to take screenshot at any occasion.
On the apk side, the obfuscation has been applied very strongly onto both class names and variable names, which makes it difficult to observe the entire protocol by just reading the decompiled Java code. 
It is cleared that the class named \texttt{ScreenshotService} is in charge of the communication with the proxy but the code is not as readable as the app I. 
What worse is that there is magic number, \texttt{89234820}, being found and referenced multiple times by reading through the assembly code of \texttt{ScreenshotService} class. 
There is a great possibility that the app has 
(1) multiple communication session with proxy to take a screenshot; and/or 
(2) an authentication trick to indicate the app's identity, which might be the reason of the existence of the magic number  \texttt{89234820}.

\subsubsection{Dynamic analysis}~\\\vspace{-8pt}

Unlike what we have done in case study of app I, only static analysis is not adequate to find the protocol that used for communication between the app II ``\emph{No Root Screentshot It}'' and its proxy. In order to unveil what kind of command that the app sends to the proxy to take screenshot and how they interact with each other, a dynamic analysis technique called ``hooking'' is adopted in this project.
Hooking system APIs on Android could be enabled by using a framework known as \emph{Xposed} on a rooted device. 
By reading the decompiled code during the static analysis, the communication between the proxy and app has been found carry out through the socket channel. 
Therefore, the monitoring of the socket channel during the communication between the proxy and app could be done by writing a module based on Xposed framework which hooks all the socket channel related packet data IO functions in specific source codes.

\begin{figure}[!tbp]
	\centering
	\begin{minipage}[t][][b]{0.42\textwidth}
		\centering
		\includegraphics[width=0.65\linewidth]{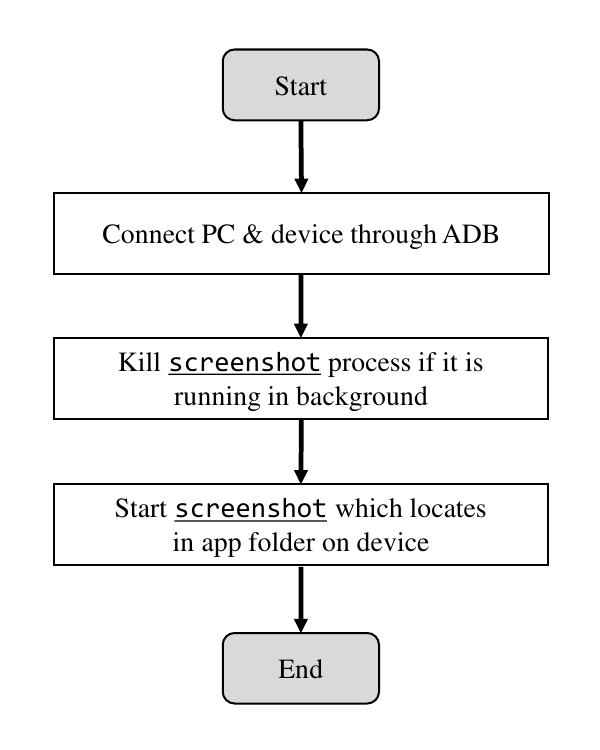}
		\caption{Proxy activation of app II}
		\label{fig:serviceActivationEdward}
	\end{minipage}
	\hfill
	\begin{minipage}[t][][b]{0.57\textwidth}
		\centering
		\vspace{-10px}
		\includegraphics[width=0.95\linewidth]{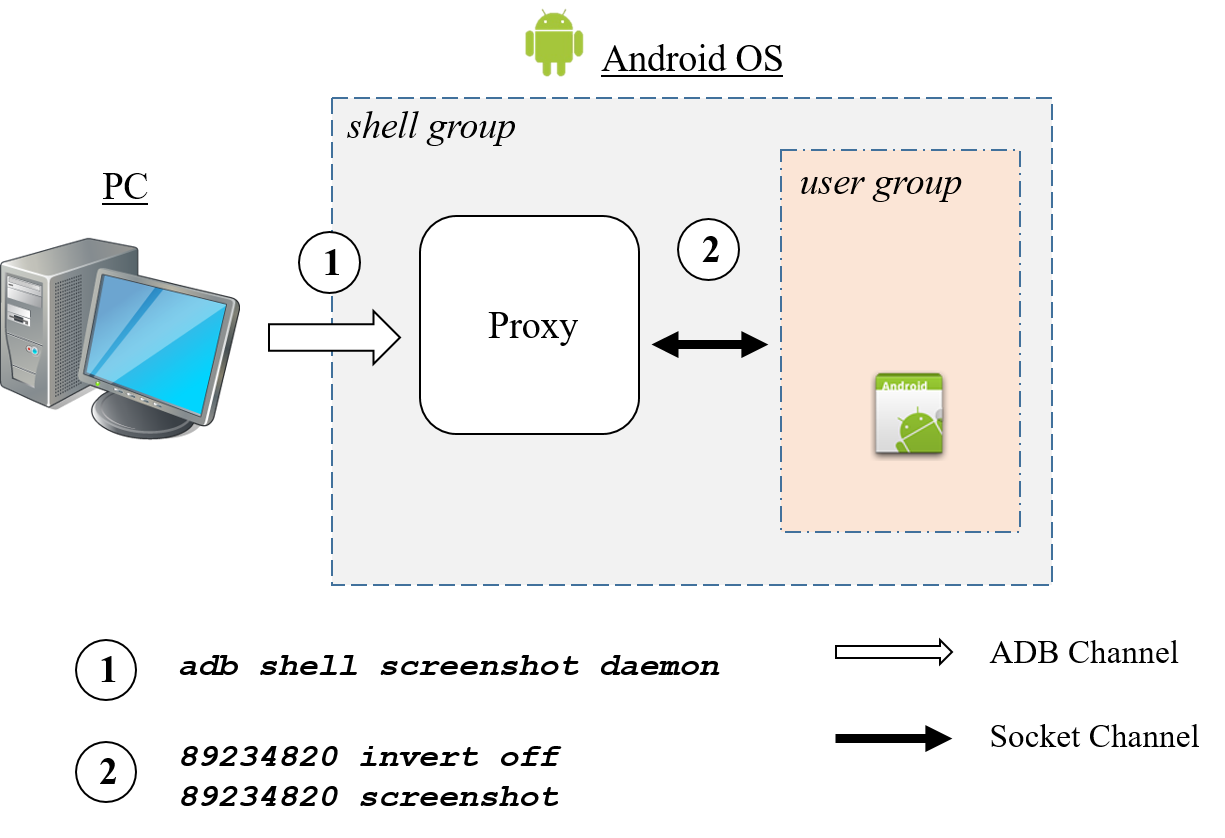}
		\caption{Process of taking screenshot on app II}
		\label{fig:processDiagramEdward}
	\end{minipage}
	\vspace{-15px}
\end{figure}

After implementing and deploying our Xposed module named ``\emph{hookNoRoot- ScreenshotIt}'', all the necessary data has been logged and printed out during the IO operation of the socket channel. The control flow of the app and proxy communication is finally unveiled and shown in Fig.~\ref{fig:processDiagramEdward}.

\subsection{App III -- ``FREE screen recorder NO ROOT''}

In addition to those 2 screenshot apps, a screen recording app ``\emph{FREE screen recorder NO ROOT}'' has also been investigated in this paper. 
According to the description on Google Play Store, this app could enable users to record their screen regardless of which app or activity is on the top of stack, and then export the recorded video in MP4 format. The entire process doesn't request users to root their devices.

\subsubsection{Analysis on the proxy activation}~\\\vspace{-8pt}

Similar with those two screenshot apps we have analyzed previously, this app also needs user to complete the proxy activation before the app unlocking the record function. 
The activation process is launched by an \emph{exe} file on Windows OS, and an executable \emph{jar} file on Linux OS.
Missing of activation script doesn't mean the identification of proxy is impossible. Actually, with the help of ADB, we could still find the details proxy(s) activated, including the process name, PID and the port(s) listening. Here, two ADB commands, ``\texttt{ps}'' and ``\texttt{netstat}'', have been used to retrieve the list of running processes and active ports on the Android device. By this means the proxy and the ports number could be found. There are two services namely ``\texttt{videoserv}'' and ``\texttt{inputserv}'' running on the background to enable users to record their screen. One of them uses port 7938, and the other one uses port 7940 to communicate with app.
However, we are still yet to completely discover the protocol without knowing the identification of the process which actively engages with those two ports. 
For that reason, the decompilation is needed for the further analysis. 

\subsubsection{Analysis on the apk file}~\\\vspace{-8pt}

Firstly, the apk file has been extracted out of the device, and then been decompiled into smali code. The clues that recently found, two port numbers 7938 and 7940, could be searched within the source code to locate the key classes we will analyze on. 
As the searching result of keyword 7938 shown below, we found the variable name which bearing that port number, namely ``\texttt{video\textunderscore port}''. Similarly, the port 7940 has been found in variable name ``\texttt{input\textunderscore port}'' recorded in same \emph{xml} file.
Next, we continued searching the occurrence of two variables ``\texttt{video\textunderscore port}'' and ``\texttt{audio\textunderscore port}''. After filtering from the search result, we preliminarily confirmed that the code reflecting the control flow and data flow was located in class ``\texttt{RecordService}'' and ``\texttt{Projection}'' separately. 
Take the communication between video server and app as example, the core function in charge of the communication flow is supposed to be ``\texttt{videoWrite}'', which located in line 1038 in the smali code of \texttt{RecordService}. This \texttt{videoWrite} method has been called many time once after the occurrence of the constant string with all letter being capitalized, which is suspected to be the command sending to the server. Moreover, by browsing through the smali code, a method named ``\texttt{openSocket}'' has been called within the class \texttt{RecordService}, which helps us to confirm that the protocol we are going to discover is performed through the socket channel.

\subsubsection{Dynamic Analysis \& authentication analysis}~\\\vspace{-8pt}

Similar with the analysis of app II, we used hooking to sketch out the complete control flow and data flow of the protocol. The target function to be hooked has been confirmed during the previous analysis, which is ``\texttt{videoWrite}'' located in class named ``\texttt{RecordService}''. In order to find as many details about the protocol as possible, some other methods located in the same class of ``\texttt{videoWrite}'' have also been hooked. 
With the information obtained from the output logs of methods' hooking, the protocol of the communication between video server and app to start screen recording has been found.
A sixteen-digit-long string \texttt{ce2757a06d455af2} grabbed our attentions because it was presumed to be the authentication code, or password for short, according to the location of its occurrence. Nevertheless, it has not yet been confirmed to be a string constant or a dynamic changing string so far. In order to clarify the nature of that password, a series of experiments has been conducted. 

Firstly, we closed the app after taking a screen recording video clips, then re-opened it and took another screen recording. 
Hooking logs shown in logcat console showed that the password didn't change. 
In that means, the password is independent of the app's life-cycle. We have repeated the above steps for many times and all the results proved that our assumption is correct.
Next, we killed all proxies related to this app and launch the service activation again. 
As a result, we noticed that the password has changed.
Thus the password could be confirmed to be dynamically generated after each time that the proxy being re-activated.
Since the password is proved to be generated by proxy, there must be a place that the proxy stored the code at somewhere that the app with user privilege could access and read. 
After searching, we finally located the password in a \emph{log} file named \texttt{videoserv.log} under the directory \texttt{data/local/tmp}, and luckily find the first occurrence of the current password was always after the word ``\texttt{AUTH}'' in the log file. 
With those information, An attacker could all along hold the current password by writing a simple program on the target device to read the log file and then extract the string at given location.

\begin{figure*}[t!]
	\vspace{-5px}
	\begin{center}
		\includegraphics[width=0.95\linewidth]{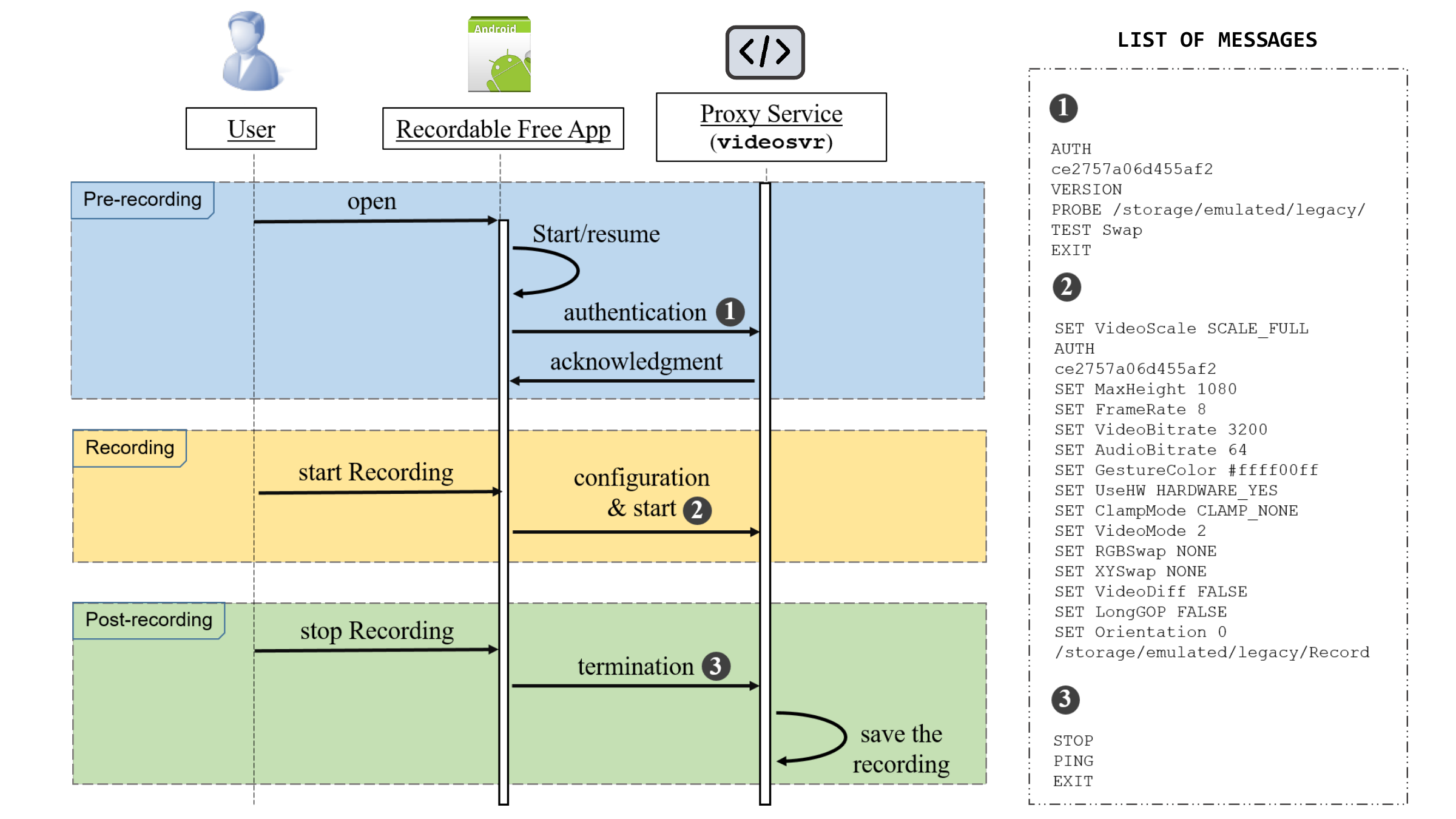}
	\end{center}
	\vspace{-15pt}
	\caption{The sequence diagram of screen video recording on app III}
	\label{fig:processRecordableFree}
	\vspace{-12pt}
\end{figure*}

\section{Mitigation}
\label{sec:mitigation}
In order to find a solution to the concerns we raised in Section~\ref{sec:related_work}, we summarize some suggestions for the Android development throughout the study and research in this project. 
The Android developers are strongly advised to raise security awareness and take some security practices into account when implementing the functionality based on ADB workaround, including:

\begin{enumerate} 
	\item \textbf{Identity verification for the application}. 
	One possible solution for this issue may be writing a \emph{handshake} process in the proxy implementation, to make both the app and the proxy exchange their authentication. And the ADB proxy will execute the command only after a successful validation. 
	Thus the proxy service can only accept the command sent from the exactly same app. Once the app is removed and re-installed, regardless of genuine app or malicious app, another handshake validation should be required thereby to ensure the ADB proxy would not be misused.
	\vspace{5pt}
	\item \textbf{Password protection for socket channel communication}. 
	Another possible solution is to implement a stronger password mechanism. 
	For the purpose to prevent from the replay attack, the password could be dynamically generated at first and then updated after a specific time period. 
	Besides that, an \emph{out-of-bounds} password mechanism could be another option as the password is randomly generated and issued by the activation program. 
	To synchronize the password between the proxy and the app, the activation program can display its password on the PC screen and ask user to manually type in the benign app. 
	Thus no other application could attain the password and thereby leaves no chance to attackers to carry out exploitation.
\end{enumerate}

\section{Conclusion}
\label{sec:conclusion}
In this paper, we analyze the approach to find privacy loophole and security vulnerabilities of ADB workaround on Android platform. 
By conducting investigation on 3 different apps, we find that most of apps using ADB workaround have risk of being exploited. 
We propose a methodology to conduct exploitation to all similar apps that use ADB workaround through the socket channel.
We also implement an exploit application on Android, which successfully proves our findings and verifies our proposed methodology.
In the end, we provide our recommendation to app developers to mitigate security risks and produce their apps with higher privacy-preserving capability.

\bibliographystyle{splncs04}
\bibliography{reference_springer} 

\end{document}